\documentclass[prd,preprintnumbers,amsmath,amssymb]{revtex4}

\begin{document}
\draft

\title{Information Loss in Black Hole Evaporation}
\author{Jingyi Zhang,\footnote{E-mail: physicz@tom.com} Yapeng Hu, Zheng Zhao\footnote{{E-mail:
zhaoz43@hotmail.com}}} \affiliation{ Department of Physics ,
Beijing Normal University, Beijing 100875, China}
\date{\today}

\begin{abstract}
Parikh-Wilczek tunnelling framework is investigated again. We
argue that Parikh-Wilczek's treatment, which satisfies the first
law of black hole thermodynamics and consists with an underlying
unitary theory, is only suitable for a reversible process. Because
of the negative heat capacity, an evaporating black hole is a
highly unstable system. That is, the factual emission process is
irreversible, the unitary theory will not be satisfied and the
information loss is possible.\\\\PACS number(s):
04.70.Dy
\end{abstract}
\maketitle \vskip2pc

In 1975, Stephen Hawking Published his astounding discovery that
black holes radiate thermally. With the emission of Hawking
radiation, the black hole could lose energy, shrink, and
eventually evaporate completely. Nevertheless, it sets up a
disturbing and difficult problem: what happens to information
during the black hole evaporation? About this puzzle, there are
two opposite opinions. Hawking and Kip Thorne thought that
information is lost in black hole evaporation. But John Preskill
believes that information is not lost and can get out of a black
hole. In 2004, Hawking changed his opinion. He thinks that
information can get out of a black hole\cite{Hawking}. In 2000,
Parikh and Wilczek investigated the black hole radiation again.
They treat Hawking radiation as a tunnelling process, and give a
semiclassical but the first explicit calculation about this
problem\cite{Parikh1}. Their result is considered to be in
agreement with an underlying unitary theory and a support to the
information conservation. Following this method, a number of
static or stationary rotating black holes are studied
\cite{Parikh2,Hemming,Medved,Alves,Vagenas1,Vagenas2,Vagenas3,Vagenas4,Vagenas5,Vagenas6,Vagenas7,Vagenas8,Zhang1,Zhang2,Zhang3,Zhang4}.
The same result, that is, Hawking radiation is no longer pure
thermal, unitary theory is satisfied and information is conserved,
is obtained. Is information conserved during the process of black
hole radiation? Does the Parikh-Wilczek tunnelling framework prove
the information conservation?  In this letter, we argue that
Parikh-Wilczek tunnelling framework is only suitable for the
reversible process. Considering the factual emission process is an
irreversible process, information loss is possible.

We first investigate the tunnelling process from the Schwarzschild
black hole. According to Ref. \cite{Parikh1}, the imaginary part
of the action for the classically forbidden trajectory is
\begin{equation}
\mathrm{Im}S=\mathrm{Im}\int^{r_f}_{r_i}p_r\mathrm{d}r=\int^{M-\omega}_M-2\pi
r'_H\mathrm{d}M'=-\frac{1}{2}[4\pi(M-\omega)^2-4\pi
M^2]=-\frac{1}{2}(S_f-S_i)\label{S1},
\end{equation}
and the emission rate is
therefore\cite{Parikh1,Keski-Vakkuri:1996xp}
\begin{equation}
\Gamma\sim e^{-2\mathrm{Im}S}=e^{\bigtriangleup S_{BH}}\label{a},
\end{equation}
which is in agreement with an unitary result in Quantum Mechanics,
$\Gamma(i\to f)=\mid M_{fi}\mid ^2\cdot e^{\bigtriangleup S}$.
Therefore, Parikh-Wilczek tunnelling framework is considered to be
a support to the information conservation. We rewrite equation
(\ref{S1}) as
\begin{equation}
\mathrm{Im}S=\mathrm{Im}\int^{r_f}_{r_i}p_r\mathrm{d}r=\int^{M-\omega}_M-\frac{1}{2}\cdot\frac{
\mathrm{d}M'}{T'}=-\int^{S_f}_{S_i}\frac{1}{2}\mathrm{dS'}=-\frac{1}{2}(S_f-S_i)\label{S2},
\end{equation}
where $T'=\frac{1}{8\pi M'}$ is the Hawking temperature, $S_i$ and
$S_f$ are the Bekenstein-Hawking entropy of the black hole
corresponding to the initial state and the final state,
respectively. From equation (\ref{S2}) we see that the emission
process satisfies the first law of black hole thermodynamics, that
is
\begin{equation}
\frac{ \mathrm{d}M'}{T'}=\mathrm{dS'}\label{S3}.
\end{equation}
In fact, similar relationship can also be found in the stationary
rotating black holes. Let us, for example, take into account the
tunnelling of an uncharged particle from the Kerr-Newman black
hole. The imaginary part of the action for the classically
forbidden trajectory is\cite{Zhang2}
\begin{eqnarray}
\mathrm{Im}S&=&\mathrm{Im}\int^{t_f}_{t_i}(L-p_\varphi
\dot{\varphi})\mathrm{d}t\nonumber\\&=&-\frac{1}{2}\int^{M-\omega}_M[\frac{4\pi
(M'^2+M'\sqrt{M'^2-a^2-q^2}-\frac{1}{2}q^2)}{\sqrt{M'^2-a^2-q^2}}\mathrm{d}M'-\frac{4\pi
(M'^2+M'\sqrt{M'^2-a^2-q^2}-\frac{1}{2}q^2)}{\sqrt{M'^2-a^2-q^2}}\Omega'_H\mathrm{d}J']\nonumber\\&=&-\frac{\pi}{2}[((M-\omega)+(M-\omega)\sqrt{(M-\omega)'^2-a^2-q^2}\,)^2-(M+M\sqrt{M'^2-a^2-q^2}\,)^2]\nonumber\\&=&-\frac{1}{2}(S_f-S_i)\label{S4}.
\end{eqnarray}
Similarly, we rewrite equation (\ref{S4}) as the following
\begin{eqnarray}
\mathrm{Im}S&=&-\frac{1}{2}\int^{M-\omega}_M[\frac{\mathrm{d}M'}{T'}-\frac{\Omega'_H\mathrm{d}J'}{T'}]=-\int^{S_f}_{S_i}\frac{1}{2}\mathrm{dS'}=-\frac{1}{2}(S_f-S_i)\label{S5},
\end{eqnarray}
where
\begin{equation}
T'=\frac{\sqrt{M'^2-a^2-q^2}}{4\pi
(M'^2+M'\sqrt{M'^2-a^2-q^2}-\frac{1}{2}q^2)}.
\end{equation}
Obviously, from equation (\ref{S5}) we can easily find the
relationship
\begin{equation}
\frac{\mathrm{d}M'}{T'}-\frac{\Omega'_H\mathrm{d}J'}{T'}=\mathrm{dS'}\label{f}.
\end{equation}
In fact, Eq. (\ref{S3}) and Eq. (\ref{f}) are the differential
forms of the first law of black hole thermodynamics\cite{Bardeen}.
It is an incorporation of the energy conservation law,
$\mathrm{d}M-\Omega_H\mathrm{d}J=\mathrm{d}Q$, and the second law
of thermodynamics, $\mathrm{d}S=\frac{\mathrm{d}Q}{T}$. For a
Schwarzschild black hole, there is no angular momentum, the first
law of the black hole thermodynamics is $dM=dQ$. The equation of
energy conservation is suitable for any process (reversible or
irreversible process). But the equation
$\mathrm{d}S=\frac{\mathrm{d}Q}{T}$ is only true for a reversible
process. For an irreversible process,
$\mathrm{d}S>\frac{\mathrm{d}Q}{T}$. That is, Parikh-Wilczek
tunnelling framework has treated the emission process as an
reversible process. In their treatment, the black hole and the
outside approach an thermal equilibrium. There is an entropy flux
$\mathrm{d}S=\frac{\mathrm{d}Q}{T}$ between the black hole and the
outside. As the black hole radiate, the entropy of the black hole
decreases, but the total entropy of the black hole and the outside
is constant, and therefore the information is conserved. Thus, in
Parikh-Wilczek tunnelling framework the result is consistent with
an unitary theory. But, in fact, because of the negative heat
capacity, an evaporating black hole (a Schwarzschild black hole or
a Kerr-Newman black hole) is (when is isolation) a highly unstable
system. With the radiation, the thermal equilibrium between the
black hole and the outside is unstable. There will be a difference
in temperature, that is, the process is irreversible. In this
case, we can not get the relationship
$\mathrm{d}S=\frac{\mathrm{d}Q}{T}$. Then the equations (\ref{S1})
and (\ref{S4}) will not be satisfied. That is, the emission rate
from the Schwarzschild black hole or from the Kerr-Newman black
hole will not be in agreement with the unitary theory. The total
entropy will increase, and therefore information is lost during
the process of the black hole radiation. That is to say,
Parikh-Wilczek tunnelling framework has not proved the information
conservation.

In fact, there are two clues for us to comprehend the information
paradox. One clue is that in physics there are many conservation
laws, such as the energy conservation law, the momentum
conservation law and the electric charge conservation law, but
there is no information conservation law. On the contrary, if the
viewpoint in Information Theory, which treats information as
negative entropy, is correct, and if the information entropy and
the thermodynamical entropy are the same in essence, then, the
information loss is a natural result. This is because of the
second law of thermodynamics--the principle of entropy increase
for an irreversible process. Now that the entropy is not
conserved, of course, information is not conserved either.

Another clue is that the treatments before in black hole
radiation, as pointed out by Hawking in 2004\cite{Hawking2}, are
too idealization. Hawking radiation, in practice, may deviate from
the perfect thermal spectrum, and maybe some ``slag" is left in
the final state of the black hole evaporation. In conclusion, a
factual Hawking process does not conserved in information. But
part of the information can get out of the black hole or stay at
the final state of the black hole.

\acknowledgments This research is supported by National Natural
Science Foundation of China (Grant No. 10373003) and National
Basic Research Program of China (Grant No. 2003CB716300).

\end{document}